\documentclass{jps-cp}

\title{Recent Advances and Precision Measurements with the JYFLTRAP Double Penning Trap Mass Spectrometer}
\author{Zhuang~\textsc{Ge}$^{1}$, Natalia~\textsc{Ambrosio-Macias}$^{1}$, Anita~\textsc{Candiello}$^{1,2}$, Saikumar~\textsc{Chinthakayala}$^{1,3}$,
Tommi~\textsc{Eronen}$^{1}$, Arthur ~\textsc{Jaries}$^{1,4,5}$, Anu~\textsc{Kankainen}$^{1}$, Angelos~\textsc{Karadimas}$^{1,2}$, Brian~\textsc{Kootte}$^{1}$, Maxime~\textsc{Mougeot}$^{1}$,
Iain~\textsc{Moore}$^{1}$, Luis~\textsc{Miguel Motilla Martinez}$^{1,3}$, Heikki~\textsc{Penttil\"a}$^{1}$, Stefanos~\textsc{Pelonis}$^{1,2}$, Prince Shaheen~\textsc{Parvez}$^{1,6}$, 
Mikael~\textsc{Reponen}$^{1}$, Simon~\textsc{Rausch}$^{1}$, Jouni~\textsc{Ruotsalainen}$^{1,6}$, Alexandre~\textsc{Roger}$^{1}$, ~\textsc{Wirunchana Rattanasakuldilok}$^{1}$, Jan~\textsc{Saren}$^{1}$,  Marek ~\textsc{Stryjczyk}$^{1,7}$, 
Jouni~\textsc{Suhonen}$^{1}$, Ville~\textsc{Virtanen}$^{1,3}$, Miikka~\textsc{Winter}$^{1}$,
and the \textsc{IGISOL Collaboration}$^{1}$}
\inst{$^{1}$Department of Physics, University of Jyv\"askyl\"a, P.O. Box 35, FI-40014 Jyv\"askyl\"a, Finland \\
$^{2}$KU Leuven, Instituut voor Kern- en Stralingsfysica, B-3001 Leuven, Belgium \\
$^{3}$GANIL, CEA/DSM-CNRS/IN2P3, Bd Henri Becquerel, 14000 Caen, France \\
$^{4}$Max-Planck-Institut für Kernphysik, Saupfercheckweg 1, 69117 Heidelberg, Germany \\
$^{5}$Experimental Physics Department, CERN, CH-1211 Geneva 23, Switzerland \\
$^{6}$GSI Helmholtzzentrum f\"ur Schwerionenforschung GmbH, 64291 Darmstadt, Germany \\
$^{7}$Institut Laue-Langevin, 38042 Grenoble, France \\
Corresponding author: \texttt{zhuang.z.ge@jyu.fi}}
\recdate{September 25, 2026}

\abst{High-precision atomic-mass measurements with the JYFLTRAP double Penning-trap mass spectrometer at the IGISOL facility provide direct information on decay energies, nuclear binding and shell evolution, isomer excitation energies, and masses relevant to nuclear astrophysics. Recent work has combined the phase-imaging ion-cyclotron-resonance (PI-ICR) method with established ion-purification and selective ion-production techniques to study low-yield nuclei and resolve isomeric states. At JYFLTRAP/IGISOL, mass-selective buffer-gas cooling, Ramsey cleaning, resonance ionization spectroscopy (RIS), phase-dependent cleaning, and species-resolved PI-ICR analysis are used in complementary combinations according to the ion yield and the required separation. Laser-frequency tagging correlates individual ion events with the corresponding laser frequency and enables state-resolved resonance-ionization spectra. Selected ion bunches are used for direct mass measurements with PI-ICR or time-of-flight ion-cyclotron resonance (TOF-ICR), for ion counting, or for post-trap decay spectroscopy. When multiple species or states are present, resolved PI-ICR phase spots allow the mass-measurement data to be assigned to the corresponding component. Examples from the recent program include high-precision $Q$-value measurements, neutron-deficient nuclei below $^{100}$Sn, neutron-rich fission fragments, rapid structural evolution around $N\approx60$, and rare-earth nuclei relevant to the astrophysical $r$ process.}

\kword{Penning-trap mass spectrometry, JYFLTRAP, PI-ICR, decay $Q$ value, nuclear isomer, low-$Q$ decay, $r$ process}

\newcommand{\jyfl}{JYFLTRAP}
\newcommand{\igisol}{IGISOL}
\newcommand{\piicr}{PI-ICR}

\begin{document}
\maketitle

\section{Introduction}

Atomic masses provide direct access to binding energies, nucleon separation energies, shell gaps, reaction thresholds, isomer excitation energies, and decay-energy releases. Precision mass measurements therefore connect nuclear-structure and astrophysics studies with investigations of weak decays and other precision applications. High precision must be accompanied by reliable isotope and nuclear-state assignment: if a low-lying isomer or another component is not resolved, the measured mass can be assigned to the wrong nuclear state, with an energy difference far larger than the statistical uncertainty of the measurement.

The \igisol{} method is particularly effective for precision mass measurements because reaction products are stopped in helium and extracted largely independently of their chemical properties~\cite{Moore2014}. At \jyfl{}, the double Penning trap combines this production mechanism with high-resolution ion purification and precision mass determination in a 7-T superconducting magnet~\cite{Eronen2012,Kankainen2020}. The mass of the ion of interest is determined from its cyclotron-frequency ratio to a reference ion. The commissioning and subsequent routine use of \piicr{} at \jyfl{}~\cite{Nesterenko2018,Nesterenko2021Phase} have been especially important because ion species and isomeric states with sufficiently different cyclotron frequencies can be resolved as distinct phase spots for suitable phase-accumulation times. Together with selective ion-source and cleaning methods, PI-ICR has enabled precision measurements of low-yield nuclei and low-lying isomeric states.

\begin{figure}[!hpt]
\centering
\includegraphics[width=0.9\textwidth]{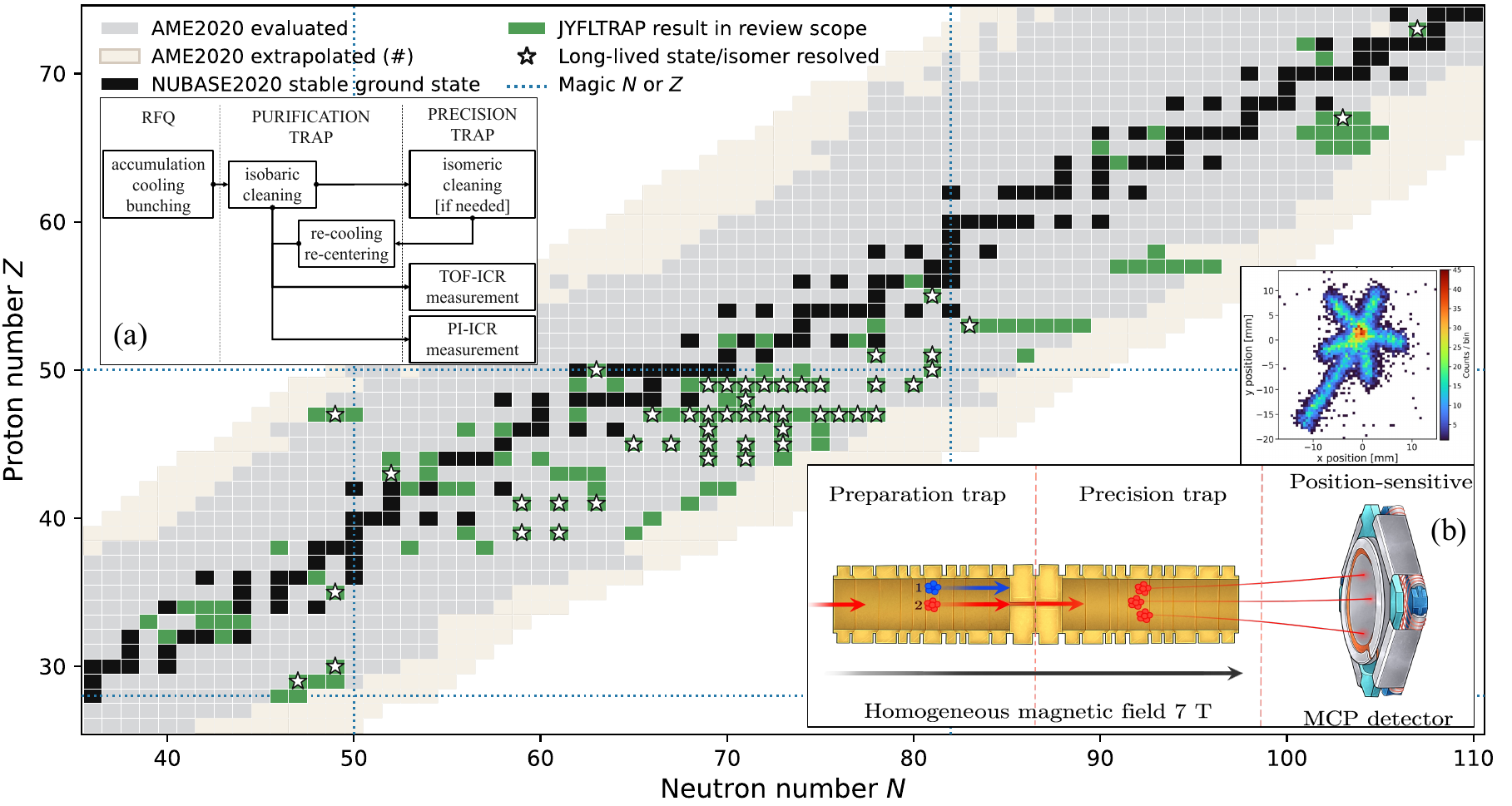}
\caption{Overview of selected recent JYFLTRAP mass, isomer, and decay-energy measurements from the 2021--2026 period. The chart contains 133 highlighted nuclides from recent JYFLTRAP measurements, including 48 for which long-lived nuclear states were resolved and it illustrates the broad coverage of the program. The upper-left flow diagram (a) summarizes RFQ accumulation, preparation-trap purification, optional isomeric cleaning, and precision-trap TOF-ICR or PI-ICR measurements. The lower-right schematic (b) shows the JYFLTRAP double-trap geometry in the 7-T magnetic field and the position-sensitive MCP detector used for phase imaging; the small inset illustrates representative ion-motion and phase-manipulation patterns used for ion centering and damping, as well as for checks of the trapping-field homogeneity.
}
\label{fig:overview}
\end{figure}

Figure~\ref{fig:overview} summarizes a selection of published JYFLTRAP results since AME2020 (2021--2026)~\cite{AME2020,NUBASE2020} discussed in this review. The highlighted nuclei span low-$Q$ $\beta$ and electron-capture decays, neutron-deficient nuclei below $^{100}$Sn, neutron-rich systems toward $^{78}$Ni and $^{132}$Sn, the rapidly changing $N\approx60$ region, and rare-earth nuclei relevant to the astrophysical $r$ process. For direct decay-energy measurements, both members of a directly measured parent--daughter pair are included. For other mass measurements, both the ion of interest and the reference nuclide are included when they are explicitly documented in the source data; this criterion is applied consistently throughout the chart.

\section{Ion preparation, selective production, and PI-ICR identification at JYFLTRAP}

The IGISOL--JYFLTRAP measurement chain combines ion production with RFQ accumulation, cooling, and bunching~\cite{Nieminen2001,Virtanen2025RFQ}, preparation-trap purification, and precision-trap mass measurements~\cite{Eronen2012,Kankainen2020}. In the precision trap, the mass is obtained from the cyclotron-frequency ratio between the ion of interest and a reference ion. PI-ICR adds position-sensitive phase imaging~\cite{Nesterenko2018,Nesterenko2021Phase}: with a suitable phase-accumulation time, ion species or nuclear states with sufficiently different cyclotron frequencies appear as separate phase spots. Selective ion production and higher-resolution cleaning can therefore be combined with PI-ICR to monitor the ion composition and assign the measured component to a specific isotope or isomeric state.

For precision mass-ratio measurements, reference-ion and ion-of-interest measurements are interleaved so that temporal magnetic-field variations can be interpolated and their effect minimized. Same-$A$ or nearby-mass reference ions are used when available to reduce mass-dependent systematic effects. Low ion counts per bunch minimize ion--ion interactions. Cleaning frequencies and phase-accumulation times are varied when needed, and the data are checked for contamination and ion-number-dependent shifts by comparing frequency ratios for different detected ion-count classes. When more than one isotope or isomeric state is present, PI-ICR phase-spot gates allow the frequency-ratio data for each resolved component to be analyzed separately. 

\begin{figure}[t]
\centering
\includegraphics[width=0.9\textwidth]{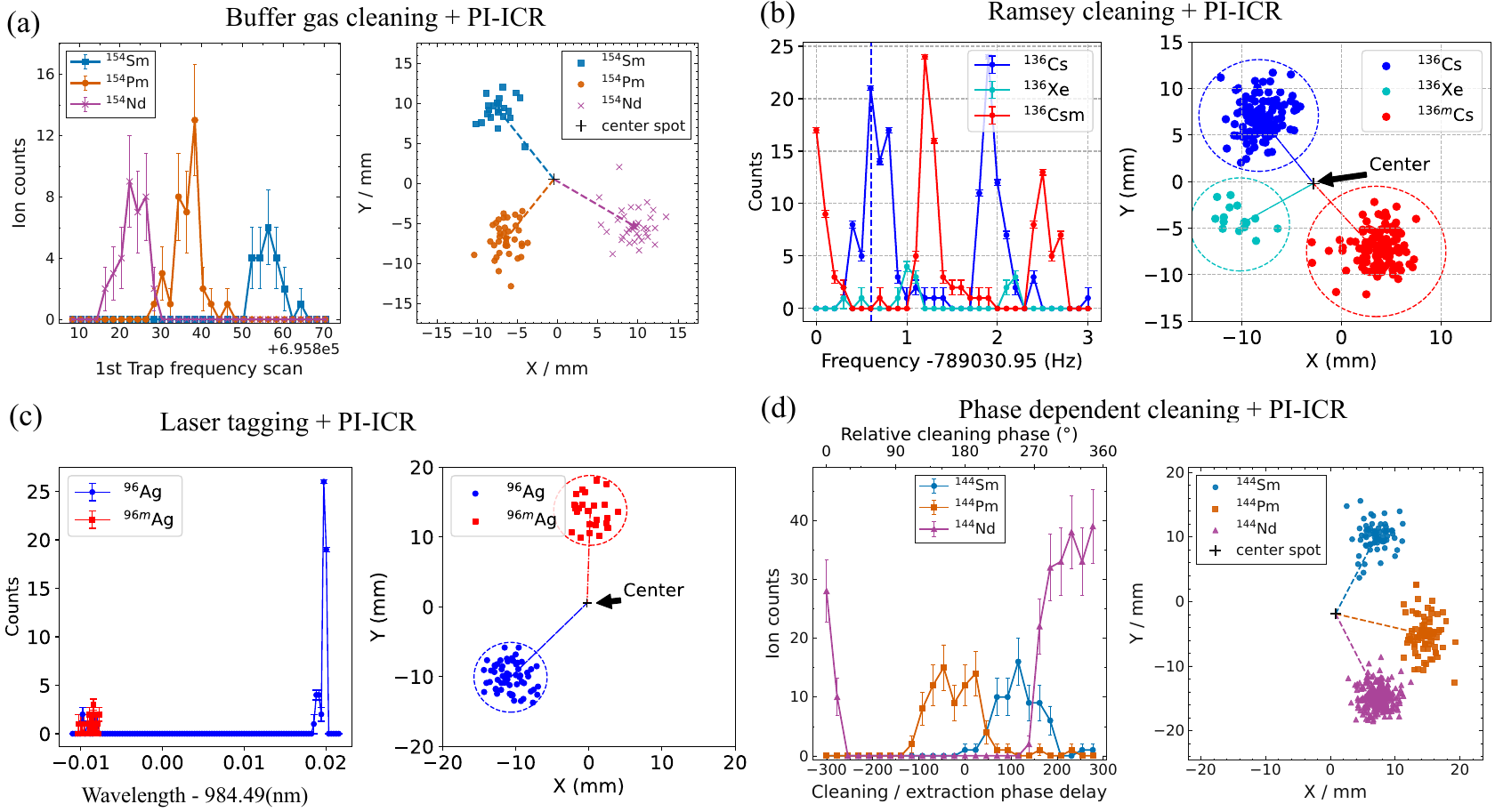}
\caption{Four representative JYFLTRAP examples of ion preparation or selective production combined with PI-ICR phase-spot identification; panels (a) and (d) show recent measurement data. (a) Preparation-trap buffer-gas cleaning of the $A=154$ isobars $^{154}$Sm, $^{154}$Pm, and $^{154}$Nd, with the corresponding PI-ICR phase spots. (b) Ramsey cleaning of the $A=136$ system, where the response is evaluated separately for $^{136}$Cs, $^{136}$Xe, and $^{136m}$Cs using PI-ICR phase-spot gates. (c) Laser-frequency tagging of $^{96}$Ag and $^{96m}$Ag; ion events selected with the corresponding phase-spot gates are correlated with the tagged laser frequency. (d) Phase-dependent separation of $^{144}$Sm, $^{144}$Pm, and $^{144}$Nd by varying the extraction-phase delay; the upper axis gives the corresponding relative cleaning phase from $0^\circ$ to $360^\circ$. The PI-ICR images allow the transmitted ion species or isomeric state to be assigned; measurements at several phase-accumulation times can be used, when required, to confirm an unambiguous assignment.}
\label{fig:cleaning}
\end{figure}

Figure~\ref{fig:cleaning} shows four examples in which ion-preparation or selective-production methods are combined with PI-ICR for isotope- or state-resolved identification.

\textbf{Buffer-gas cleaning + PI-ICR.} Mass-selective buffer-gas cooling in the preparation trap is used to purify and center the ions before transfer to the precision trap. Cleaning-frequency scans can be evaluated using species-resolved PI-ICR ion counts. In the $A=154$ example in Fig.~\ref{fig:cleaning}(a), $^{154}$Sm, $^{154}$Pm, and $^{154}$Nd are clearly separated in the corresponding PI-ICR phase image. Applying phase-spot gates makes it possible to evaluate the preparation-trap cleaning response species by species and to select the quadrupolar cleaning frequency for the desired component. The panel is included as a representative recent measurement example of the combined cleaning and PI-ICR identification procedure.

\textbf{Ramsey cleaning + PI-ICR.} When the required separation cannot be achieved with buffer-gas cleaning alone, Ramsey dipolar cleaning in the precision trap is used to separate ion species or isomeric states with small mass differences. The unwanted component is excited to a larger radial amplitude and removed during transfer back through the narrow diaphragm, 
 after which the remaining ions are re-cooled and re-centered in the preparation trap. In the $A=136$ example in Fig.~\ref{fig:cleaning}(b), the cleaning-frequency response is determined separately for $^{136}$Cs, $^{136}$Xe, and $^{136m}$Cs using PI-ICR phase-spot gates for the corresponding resolved components; the same phase image verifies the ion composition after cleaning. This procedure was applied in the direct $^{136}$Cs--$^{136}$Ba decay-energy measurement~\cite{Cs136}.

\textbf{Laser-frequency tagging + PI-ICR.} Resonance ionization spectroscopy provides element and/or state selectivity at the ion source. When the used laser frequency is recorded for each detected ion, events from a chosen isotope or isomeric state can be selected with PI-ICR phase-spot gates; correlating those events with the tagged laser frequency then yields a separate laser spectrum for that component. Figure~\ref{fig:cleaning}(c) illustrates this procedure for $^{96}$Ag and $^{96m}$Ag. PI-ICR-assisted resonance-ionization spectroscopy was previously demonstrated for neutron-deficient silver isotopes at very low ion rates~\cite{Reponen2021AgRIS}. The hot-cavity catcher laser ion source together with PI-ICR was subsequently used for the high-precision mass measurements of $^{95-97}$Ag below $^{100}$Sn~\cite{Ag9597}.

\textbf{Phase-dependent cleaning + PI-ICR.} Close-lying ion species can also be separated using the difference in their accumulated radial phases. After a selected phase-evolution time,  the extraction timing for transfer back through the narrow diaphragm to the preparation trap is adjusted so that the desired component is transmitted while another component is suppressed. The lower-right panel of Fig.~\ref{fig:cleaning} illustrates this procedure for $^{144}$Sm, $^{144}$Pm, and $^{144}$Nd: the relative ion populations change with the extraction-phase delay, and the transmitted species are resolved in the PI-ICR image. Phase-based separation was demonstrated at \jyfl{} with short-lived $^{127}$In states~\cite{Nesterenko2018} and provides an additional option when species with small mass differences must be separated within a short preparation cycle.

These techniques provide complementary selectivity at different stages of the experimental cycle: resonance ionization at production, mass-selective buffer-gas cooling in the preparation trap, Ramsey or phase-dependent cleaning when higher resolving power is required, and PI-ICR phase-spot identification during the precision measurement. In laser-spectroscopy measurements, ion events selected by the phase-spot gates can also be correlated with the tagged laser frequency. The combined approach has enabled precision studies of low-yield isotopes, including state-resolved mass measurements, direct decay-energy measurements, and resonance-ionization-spectroscopy studies.


JYFLTRAP is also being coupled to complementary IGISOL devices. RAPTOR was developed to combine resonance-ionization spectroscopy with ion manipulation and laser-assisted state-selective purification in conjunction with JYFLTRAP~\cite{Kujanpaa2023}. The commissioned IGISOL MR-TOF mass spectrometer provides rapid mass analysis, separation, and ion counting~\cite{Virtanen2026MRTOF}. These developments complement the established JYFLTRAP preparation sequence rather than forming part of every measurement cycle.

Many of the observables discussed below are obtained from mass differences. One- and two-neutron separation energies trace changes in nuclear binding, a ground state--isomer mass difference gives the corresponding excitation energy, and a parent--daughter mass difference gives the decay $Q$ value. In Penning-trap measurements these mass differences are obtained from cyclotron-frequency ratios. With PI-ICR and state-selective preparation, the measured component can be assigned to the corresponding isotope or isomeric state; in suitable cases, relative populations can also be determined. Across these measurements, the preparation sequence is chosen to preserve sufficient yield while removing or identifying components that could bias the mass determination. The optimum combination therefore depends on the ion yield, half-life, and isomeric composition, with PI-ICR phase-spot gates providing the final state-resolved event assignment.

\section{Low-$Q$ decays and neutrino-mass candidates}

A low-$Q$ $\beta$-decay or electron-capture branch energy enhances the relative spectral weight near the corresponding endpoint, but a small ground-state-to-ground-state mass difference alone does not define a useful neutrino-mass candidate. The final-state excitation energy, transition type, and partial half-life must also be known; for electron capture, atomic-vacancy energies and widths, shell occupancy, electron density at the nucleus, exchange and overlap corrections, and shake-up and shake-off effects also enter. The recent \jyfl{} program therefore uses direct mass differences as the experimental energy anchor and combines them with nuclear spectroscopy and atomic calculations.

The measurements have both confirmed and excluded proposed low-$Q$ branches. For $^{159}$Dy, the directly measured $Q_{\rm EC}=364.73(19)$ keV gives branch energies of 1.18(19) and 2.68(19) keV to selected $^{159}$Tb states~\cite{Dy159}. The $^{131}$I measurement gives $Q_{\beta^-}=972.25(19)$ keV and establishes an allowed branch with $Q^*=1.03(23)$ keV~\cite{I131}. In contrast, improved masses of $^{72}$As, $^{76}$As, and $^{155}$Tb showed that several previously proposed ultralow transitions are energetically forbidden~\cite{As72,As76Tb155}. The $^{136}$Cs measurement also changed the interpretation of two proposed low-energy branches: one becomes energetically forbidden, while another remains allowed with a few-keV decay energy~\cite{Cs136}.

A wider set of measurements has progressively replaced evaluation-based branch energies by direct parent--daughter mass differences. Measurements involving $^{75}$Se/$^{75}$As constrained an ultralow electron-capture branch in the $A=75$ system~\cite{As75}. For $^{111}$In, the measured $Q_{\rm EC}=857.63(17)$ keV leads to candidate branches in the few-keV range when combined with the $^{111}$Cd level scheme~\cite{In111}. These cases show the different possible outcomes of a low-$Q$ search: a candidate can survive with a substantially reduced energy uncertainty, move outside the interesting range, or be excluded once the branch energy is fixed.

Recent technetium measurements demonstrate why state resolution and decay-energy precision must be treated together. For $^{95}$Tc, the 38.910(40)-keV isomer had to be separated from the ground state before a reliable electron-capture $Q$ value could be obtained~\cite{Tc95}. For $^{97}$Tc, the directly measured $Q_{\rm EC}=324.82(21)$ keV improves the AME2020-based precision by about a factor of 19 and gives $Q^*_{\rm EC}=4.8(10)$ keV for the candidate transition to the 320(1)-keV state in $^{97}$Mo~\cite{Tc97}. Similarly, the $^{77}$As--$^{77}$Se mass difference improved the decay-energy precision by about a factor of 24 relative to AME2020 and fixes a $Q^*=4.360(70)$ keV branch~\cite{As77}.

Two cases further extend the accessible energy scale. For $^{110m}$Ag, a high-precision $^{109}$Ag mass measurement using $^{110}$Cd as reference, combined with known spectroscopic information, gives an allowed $\beta$ branch with $Q^*=405(135)$ eV~\cite{Ag110m}. For $^{113}$Sn, the 77.389(19)-keV isomer was resolved and the ground-state electron-capture energy was determined as 1039.25(19) keV; an allowed 9.60(20)-keV branch lies near the $L$-shell binding energies of the daughter atom, where many-electron atomic effects become important~\cite{Sn113}. These measurements illustrate the main role of Penning-trap data in neutrino-mass candidate searches: not simply to find the smallest tabulated $Q$ value, but to determine the relevant branch energy with a well-defined nuclear-state assignment.

\section{Masses for nuclear structure, astrophysics, and precision applications}

Below $^{100}$Sn, the masses of $^{95-97}$Ag and the resolved long-lived states in $^{96}$Ag provide a direct test of the $N=50$ shell closure on the proton-rich side~\cite{Ag9597}. The $^{95}$Ag mass was determined directly for the first time, while the long-lived states in $^{96}$Ag were separated using the PI-ICR method. The resulting separation-energy trends support a robust $N=50$ shell closure and provide improved reaction energetics for calculations involving the neutron-deficient Ag chain. In the same experiment, selective hot-cavity laser ionization enhanced the production of the silver isotopes, while PI-ICR provided the precision mass measurement and separation of the long-lived states.

Toward doubly magic $^{78}$Ni, measurements of $^{74,75}$Ni and $^{76-78}$Cu extended the experimental mass surface and improved the corresponding separation-energy systematics~\cite{Ni78}. A dedicated high-resolution study of $^{76}$Cu subsequently resolved the two long-lived states and established an excitation energy of 64.8(25) keV, reversing the earlier ground-state/isomer assignment~\cite{Cu76}. The ground and isomeric states in $^{79}$Zn were also measured with high precision; the isomer excitation energy of about 942 keV provides an important benchmark for shape coexistence close to $^{78}$Ni~\cite{Zn79}.

The neutron-rich $A\approx100$ region shows how precision mass and state-resolved measurements trace rapid structural changes. Precision masses of $^{104}$Y, $^{106}$Zr, $^{104,104m,109}$Nb, and $^{111,112}$Mo mapped the mass surface across the neutron midshell at $N=66$~\cite{ZrN66}. New $^{104-106}$Tc masses remove irregularities in earlier separation-energy trends and indicate a smoother evolution through the $N\approx60$ region~\cite{Tc104106}. State-resolved measurements of $^{98,100}$Y and $^{100,102}$Nb, including states populated through in-trap $\beta$ decay, add isomeric-yield information and clarify the unusual structure of $^{98}$Y~\cite{N60_2025}. Measurements of $^{91}$Sr and $^{95}$Y provide further anchors for the same rapidly changing mass surface~\cite{SrY2024}.

A broad set of fission-fragment measurements has extended state-resolved information through neutron-rich Ru, Rh, Pd, Ag, Cd, In, Sn, Sb, and I isotopes. Long-lived states in $^{113,115}$Ru and $^{110,112,114,116,118}$Rh have been separated directly, and an additional long-lived state in $^{114}$Rh was identified in a later dedicated measurement~\cite{Ru2023,Rh2023,Rh114_2025}. The neutron-rich Ag chain $^{113-125}$Ag has been studied through complementary Penning-trap and laser-spectroscopy measurements, including the ordering of long-lived states and the discovery of a new isomer in $^{118}$Ag~\cite{AgOdd,AgEven}. In $^{124}$Ag the 188.2(25)-keV isomer excitation energy was determined directly and the improved $^{124,125}$Ag masses reduce the mass contribution to the uncertainty of the corresponding neutron-capture rate~\cite{Ag124125}.

Measurements of neutron-rich Cd and In isotopes and dedicated studies of low-lying In isomers have removed ambiguities from unresolved states and improved local separation-energy trends~\cite{CdIn2023,InIsomers2023}. A wider fission-fragment campaign determined new or improved state energies including $^{119m}$Pd and $^{132m}$Sb~\cite{FissionIsomers}. Around $^{132}$Sn, measurements of $^{137}$Sb and $^{136-142}$I extend the mass surface on the neutron-rich side and constrain reaction and decay energetics~\cite{Sn132region}. The $^{131}$Sn mass has also been used together with isomerically purified $^{131}$In decay spectroscopy to refine the level structure of $^{131}$Sn~\cite{Sn131}. The recent $^{129}$Sb measurement gives $Q_\beta=2411.3(12)$ keV and excludes the previously proposed ultralow ground-state branches. For $^{129m}$Sb, the revised mass gives $Q^*\approx3.3(13)$ keV to a known $^{129}$Te level, but the unknown spin-parity of that level and the high spin of the isomer make its relevance as an ultralow-$Q$ candidate doubtful~\cite{Sb129}.

In the rare-earth region, high-precision masses of $^{167-169}$Tb, $^{167-171}$Dy, and $^{169-171}$Ho extend the separation-energy systematics to the neutron midshell at $N=104$~\cite{RareEarth2024}. The masses of $^{169}$Tb, $^{170}$Dy, and $^{171}$Dy were determined experimentally for the first time, and the two long-lived states in $^{170}$Ho were resolved with $E_x=150.8(54)$ keV. The resulting neutron-capture rates shift calculated rare-earth $r$-process abundances by roughly 15--30\% for some trajectories relative to results obtained with AME2020 masses. Complementary JYFLTRAP measurements of $^{148-153}$La and $^{151}$Ce initially indicated a pronounced bump in the lanthanum two-neutron separation energies around $N\approx92$--94, with an energy gain of about 0.4 MeV~\cite{La2025}. A subsequent independent mass--lifetime measurement identified a lower-mass $^{149}$La ground state, removing the previously inferred bump and replacing it with a kink-like structure~\cite{La149_2026}. The precise $^{173}$Hf mass provides an additional constraint on the structure of $^{173}$Lu and on reaction energetics relevant to the $\gamma$ process~\cite{Hf173}.

Precision mass spectrometry also provides decay energies and excitation energies for near-stable systems. The $^{84}$Sr mass was measured with a relative precision of $4.8\times10^{-10}$, about 30 times more precise than the AME2020 value, giving a directly determined $2\beta^+$ decay energy of 1790.115(37) keV~\cite{Sr84}. Direct $Q_{\beta\beta}$ measurements of $^{98}$Mo and $^{104}$Ru reduce the energy-scale uncertainty entering double-beta-decay calculations~\cite{Mo98,Ru104}; for $^{104}$Ru the measured value is $Q_{\beta\beta}=1297.705(36)$ keV. More recently, the $^{122}$Sn double-beta-decay energy was determined as 373.58(12) keV, improving the precision by about a factor of 20; improved information was also obtained for $^{124}$Sn~\cite{Sn122124}. Direct ground--isomer mass differences also provide spectroscopic information independently of a decay scheme, as illustrated by the first direct determination of the $^{180m}$Ta excitation energy, 76.79(55) keV~\cite{Ta180}.



\section{Summary and outlook}

Recent JYFLTRAP work has increasingly combined established purification and selective-production methods with PI-ICR to extend precision measurements to low-yield nuclei and low-lying isomers while identifying or suppressing residual contaminants. The measurements reviewed here demonstrate how state-resolved mass measurements and direct decay-energy determinations contribute to low-$Q$ decay studies, shell-evolution and deformation systematics, astrophysical reaction energetics, and precision $Q$-value measurements.
Future developments will further broaden this approach through faster and more selective preparation of short-lived ions and stronger coupling to complementary IGISOL devices. RAPTOR~\cite{Kujanpaa2023} represented an important step toward combining resonance-ionization spectroscopy, ion manipulation, and laser-assisted state-selective purification with JYFLTRAP. A dedicated laser-cooling trap is currently under development for improved ion cooling, manipulation, and preparation. The commissioned IGISOL MR-TOF mass spectrometer~\cite{Virtanen2026MRTOF} provides rapid mass analysis, separation, and ion counting and offers an additional purification stage before precision trapping.
In parallel, stronger coupling to selective ion sources, improved state-selective and phase-dependent preparation, and downstream measurements with purified ion bunches will provide additional capabilities for ion counting and decay spectroscopy. Together, these developments preserve the core role of JYFLTRAP as a high-precision mass spectrometer while extending its capability to prepare isotope- and state-selected ion samples for complementary experiments.

\section*{Acknowledgments}

This work was supported by the Research Council of Finland (RCF) through Projects Nos. 314733, 315179, 320062, 327629, 345869, and 368322. Z.G. acknowledges support from RCF Project No. 354589. A.K. acknowledges support from RCF Projects Nos. 354968, 369714, and 374065. Support from the European Union's Horizon 2020 research and innovation programme under Grant Agreements Nos. 771036 (ERC CoG MAIDEN) and 861198 (LISA) is acknowledged. This project has also received funding from the European Union's Horizon Europe research and innovation programme under Grant Agreement No. 101057511 (EURO-LABS). J.S. acknowledges support from the NEPTUN project (PNRR-I8/C9-CF264, Contract No. 760100/23.05.2023 of the Romanian Ministry of Research, Innovation and Digitization).

\end{document}